\documentclass[aps,showpacs,superscriptaddress,nofootinbib,groupedaddress]{revtex4}

\usepackage{color}
\usepackage{graphicx}
\usepackage{amsmath}
\usepackage{amssymb}
\usepackage[colorlinks=true,citecolor=darkred,urlcolor=darkred, pdfborder={0 0 0}]{hyperref}
\definecolor{darkred}{rgb}{0.6,0,0}

\definecolor{linkcolor}{rgb}{0,0,0.5}

\def\gsim{\raise0.3ex\hbox{$\;>$\kern-0.75em\raise-1.1ex\hbox{$\sim\;$}}}
\def\lsim{\raise0.3ex\hbox{$\;<$\kern-0.75em\raise-1.1ex\hbox{$\sim\;$}}}

\def\beqn#1{\begin{equation}\label{#1}}
\def\eeqn{\end{equation}}

\def\beqa#1{\begin{eqnarray}\label{#1}}
\def\eeqa{\end{eqnarray}}

\def\lr{$\mathrm{SU(3)_c \otimes SU(2)_L \otimes SU(2)_R \otimes U(1)_{B-L}}$ }
\def\3211{$\mathrm{SU(3)_c \otimes SU(2)_L \otimes U(1)_R \otimes U(1)_{B-L}}$ }
\def\321{$\mathrm{SU(3) \otimes SU(2) \otimes U(1)}$ }
\def\422{$\mathrm{SU(4) \otimes SU(2) \otimes SU(2)_R}$ }

\newcommand{\Tr}{\mathrm{Tr}}

\def\Z2{$\mathcal{Z_2}$}

\def\vev#1{\left\langle #1\right\rangle}

\newcommand{\sm}{{Standard Model }}
\newcommand{\AddrAHEP}{%
  AHEP Group, Institut de F\'{i}sica Corpuscular --
  C.S.I.C./Universitat de Val\`{e}ncia, Parc Cient\'ific de Paterna.\\
 C/ Catedr\'atico Jos\'e Beltr\'an, 2 E-46980 Paterna (Valencia) - SPAIN}
 \def\one{\ensuremath{\mathbf{1}}}
 \def\two{\ensuremath{\mathbf{2}}}
 \def\three{\ensuremath{\mathbf{3}}}
 \def\threeS{\ensuremath{\mathbf{3^*}}}	
\def\six{\ensuremath{\mathbf{6}}}

\def\TrTrOne{ $\mathrm{ SU(3)_c \otimes SU(3)_L \otimes U(1)_{X'} }$ }
\def\TrTrTrOne{$\mathrm{SU(3)_c \otimes SU(3)_L \otimes SU(3)_R \otimes U(1)_{X}}$ }
\begin{document}

 \title{Unifying left-right symmetry and 331 electroweak theories}
\author{Mario Reig}
\email{mareiglo@alumni.uv.es}
\affiliation{\AddrAHEP}
\author{Jos\'e W.F. Valle}
\email{valle@ific.uv.es}
\affiliation{\AddrAHEP}
\author{C.A. Vaquera-Araujo}
 \email{vaquera@ific.uv.es}
\affiliation{\AddrAHEP}
\date{\today}

\pacs{14.60.Pq, 12.60.Cn, 12.15.Ff, 14.60.St}

%%%%%%%%%%%%%%%%%%%

\begin{abstract}
\noindent

We propose a realistic theory based on the 
$\mathrm{SU(3)_c \otimes SU(3)_L \otimes SU(3)_R \otimes U(1)_{X}}$ gauge group
which requires the number of families to match the number of
colors. In the simplest realization neutrino masses arise from the
canonical seesaw mechanism and their smallness correlates with the
observed V-A nature of the weak force. Depending on the symmetry
breaking path to the Standard Model one recovers either a left-right symmetric
theory or one based on the $\mathrm{SU(3)_c \otimes SU(3)_L \otimes U(1)}$ 
symmetry as the ``next'' step
towards new physics.
\end{abstract}
\pacs{12.60.Cn, 11.30.Er, 12.15.Ff, 13.15.+g } 
 
\maketitle

\section{introduction}
\label{motivation}

Despite its great success, the Standard Model (SM) is an incomplete
theory, since it fails to account for some fundamental phenomena such
as the existence of neutrino masses, the underlying dynamics
responsible for their smallness, the existence three families, the
role of parity as a fundamental symmetry, as well as many other issues
associated to cosmology and the inclusion of gravity. Here we take the
first three of these shortcomings as valuable clues in determining the
next step in the route towards physics Beyond the Standard Model.

One unaesthetic feature of the \sm is that the chiral nature of the
weak interactions is put in by hand, through explicit violation of
parity at the fundamental level.
Moreover the Adler–Bell-Jackiw
anomalies~\cite{Adler:1969gk,Bell:1969ts} are canceled miraculously
and thanks to the \textit{ad-hoc} choice of hypercharge assignments.
Left-right symmetric schemes such as Pati-Salam~\cite{pati:1974yy} or
the left-right symmetric models can be made to include parity and
offer a solution to neutrino masses through seesaw
mechanism~\cite{GellMann:1980vs,yanagida:1979as,mohapatra:1980ia,Schechter:1980gr,Lazarides:1980nt}
and a way to ``understand'' hypercharge~\cite{mohapatra:1980ia}.
However in this case the number of fermion families is a free
parameter.

Conversely, \TrTrOne schemes provide an explanation to the family
number as a consequence of the quantum consistency of the
theory~\cite{Singer:1980sw,valle:1983dk}, but are manifestly chiral,
giving no dynamical meaning to parity.
Even if these models allow for many ways to understand the smallness
of neutrino mass either through radiative
corrections~\cite{Boucenna:2014ela,Boucenna:2014dia} or through
various variants of the seesaw
mechanism~\cite{Montero:2001ts,Catano:2012kw,Addazi:2016xuh,Valle:2016kyz,Reig:2016ewy},
their explicit chiral structure prevents a dynamical understanding of
parity and its possible relation to the smallness of neutrino mass,
precluding a deeper understanding of the meaning of the hypercharge
quantum number.

In this paper we will address some of these issues jointly, suggesting
that they are deeply related. Our framework will be an extended
left-right symmetric model which implies the existence of mirror gauge
bosons, i.e. in addition to weak gauge bosons we have right-handed
gauge bosons so as to restore parity at high energies.
We propose a unified description of left-right symmetry and 331
electroweak theories in terms of the extended \TrTrTrOne gauge group
as a \textit{common ancestor}: Depending on the spontaneous symmetry
breaking path towards the \sm one recovers either conventional \lr
symmetry or the $\mathrm{SU(3)_c \otimes SU(3)_L \otimes U(1)}$
symmetry as the missing link on the road to physics beyond the
\sm. Other constructions adopting these symmetries have been already
mentioned in the literature. In \cite{Dias:2010vt,Ferreira:2015wja} a
model for neutrino mass generation through dimension 5 operators is
studied, and in \cite{Huong:2016kpa,Dong:2016sat} models for the
diphoton anomaly and dark matter were presented.

This work is organized as follows. We first construct a new left-right
symmetric theory showing how the gauge structure is deeply related
both to anomaly cancellation as well as the presence of parity at the
fundamental level. In the next sections we build a minimal model where
neutrino masses naturally emerge from the seesaw mechanism. Finally we
study the symmetry breaking sector, identifying different patterns of
symmetry breaking and showing how they are realized by different
hierarchies of the relevant vacuum expectation values. In the
appendix we outline the anomaly cancellation in the model.

\section{The model}
\label{model}

In this paper we propose a class of manifest left-right symmetric models based
on the extended electroweak gauge group \TrTrTrOne in which the
electric charge generator is written in terms of the diagonal
generators of $\mathrm{SU(3)_{L,R}}$ and the $\mathrm{U(1)_{X}}$
charge as
\begin{equation}\label{elcharge}
Q=T^{3}_{L} +T^{3}_{R}+\beta (T^{8}_{L}+T^{8}_{R}) + X ,
\end{equation}
where $\beta$ is a free parameter that determines the electric charge
of the exotic fields of the model~\cite{Buras:2014yna,Fonseca:2016tbn,Fonseca:2016xsy}, and its value
is restricted by the $\mathrm{SU(3)_{L,R}}$ and $\mathrm{U(1)_{X}}$
coupling constants $g_L=g_R=g$ and $g_X$ to comply with the relation
\begin{equation}
\frac{g_X^2}{g^2}=\frac{\sin^2\theta_W}{1-2(1+\beta^2)\sin^2\theta_W},
\end{equation}
with $\theta_W$ as the electroweak mixing angle
\cite{Dias:2010vt}. This relation implies that
$\beta^2 <-1+1/(2\sin^2\theta_W)$, consistent with the original
models~\cite{Singer:1980sw,valle:1983dk}~\footnote{The model is unique
  up to exotic fermion charge assignments. Unfortunately, the choice
  $\beta =\sqrt{3}$ is excluded by consistency of the model.}  Notice
that special features may arise for specific choices, such as
$\beta=0$, which contains fractionally charged
leptons~\cite{Langacker:2011db}. In a general
$\mathrm{ SU(N)_L\otimes SU(M)_R\otimes U(1)_X}$ theory $\beta=0$
always implies that the charge $X$ becomes proportional to
$B-L$. Moreover, for $N,M> 2$ the number of families must match the
number of colors in order to cancel the anomaly, since two quark
families transform as the fundamental representation and one as
anti-fundamental.  
In order to illustrate the peculiar features of this class of models,
that combine inherent aspects of both left-right symmetric models and
331 gauge structures, we will consider throughout this work, the
general case where $\beta$ is not fixed.
\begin{table}[!h]
\begin{center}
\begin{tabular}{|c||c|c||c|c|c|c||c|c|c|c|}
\hline
 & $\psi_{aL}^\ell$ & $\psi_{aR}^\ell$  & $Q_L^{\alpha}$  & $Q_R^{\alpha}$ & $Q_L^3$ & $Q_R^3$   & $\phi$ &$\rho$& $\Delta_L$ & $\Delta_R$ \\
\hline
\hline
$\mathrm{SU(3)_c}$ & \one &\one &\three &\three &\three &\three  &\one&\one&\one &\one \\
\hline
$\mathrm{SU(3)_L}$ & \three & \one & \threeS  & \one & \three & \one  & \three &\three & \six & \one   \\
\hline
$\mathrm{SU(3)_R}$ &\one &\three &\one  &\threeS &\one &\three &\threeS &\three&\one &\six\\
\hline
$\mathrm{U(1)_{X}}$ & $\frac{q-1}{3}$ & $\frac{q-1}{3}$  & $-\frac{q}{3}$  & $-\frac{q}{3}$ & $\frac{q+1}{3}$ & $\frac{q+1}{3}$ & $0$ &$\frac{2q+1}{3}$ & $\frac{2(q-1)}{3}$ & $\frac{2(q-1)}{3}$\\ 
\hline
\hline
\end{tabular}\caption{Particle content of the model, with $a=1,2,3$ and $\alpha=1,2$. See text for the definition of $q$.} 
\label{tab:content}
\end{center}
\end{table}

The particle content and the transformation properties of the fields
are summarized in table \ref{tab:content}. We assume manifest
left-right symmetry, implemented by an additional $\mathbf{Z}_2$
symmetry that acts as parity, interchanging $\mathrm{SU(3)_L}$ and
$\mathrm{SU(3)_R}$ and transforming the fields as
$\psi^{a}_{L}\leftrightarrow\psi^{a}_{R}$,
$Q^{\alpha}_L\leftrightarrow Q^{\alpha}_R$,
$Q^{3}_L\leftrightarrow Q^{3}_R$, $\rho\leftrightarrow \rho^T$,
$\phi\leftrightarrow \phi^\dagger$ and
$\Delta_{L}\leftrightarrow\Delta_{R}$.

Fermion fields are arranged in chiral multiplets
\begin{equation}
\psi^{a}_{L,R}=\left (
\begin{array}{c}
\nu\\
\ell^{-} \\
\chi^q \\
\end{array}\right )^{a}_{L,R}\,,\qquad
Q_{L,R}^\alpha=\left (
\begin{array}{c}
d\\
-u\\
J^{-q-\frac{1}{3}}\\
\end{array}\right )^{\alpha}_{L,R}\,,\qquad
Q_{L,R}^3=\left (
\begin{array}{c}
u\\
d\\
J^{q+\frac{2}{3}}\\
\end{array}\right )^3_{L,R},
\end{equation}
transforming as triplets or antitriplets under both
$\mathrm{SU(3)_{L,R}}$ groups. The electric charge of the third
components of $\psi^{a}_{L,R}$ determined by $q$ is related to the
parameter $\beta$ through
\begin{equation}
  \label{eq:beta}
\beta =-(2q+1)/\sqrt{3}  .
\end{equation}
In this setup, the mechanism behind anomaly cancellation is analogous
to the one characterizing 331 models, as detailed in Appendix
\ref{AnomalyCancellation}. Thus the first interesting result in this
framework is the fact that quantum consistency requires that the
number of triplets must be equal to the number of antitriplets. This
can be achieved if two quark multiplets transform as triplets whereas
the third one transforms as an antitriplet, which in turn implies that
the number of generations must coincide with the number of colors, an
appealing property of 331 models~\cite{Singer:1980sw}.

The scalar sector needed for spontaneous symmetry breaking and fermion
mass generation contains a bitriplet
\begin{equation}\begin{split}
&\phi=\left(
 \begin{array}{ccc}
 \phi^{0}_{11} & \phi^{+}_{12} & \phi^{-q}_{13}\\ 
 \phi^{-}_{21} & \phi^{0}_{22} & \phi^{-q-1}_{23}\\
 \phi^{q}_{31} & \phi^{q+1}_{32} & \phi^{0}_{33}
 \end{array}\right)\sim(\mathbf{3}_\mathrm{L},\mathbf{3}^{*}_\mathrm{R})\,,
\end{split}\end{equation} 
a bi-fundamental field 
\begin{equation}\begin{split}
&
\rho=\left(\begin{array}{ccc}
\rho_{11}^{+}&\rho_{12}^{0}& \rho_{13}^{q+1}\\
\rho_{21}^{0}&\rho_{22}^{-}& \rho_{23}^{q}\\
\rho_{31}^{q+1}& \rho_{32}^{q}& \rho_{33}^{2q+1}
\end{array}\right)\sim(\mathbf{3}_\mathrm{L},\mathbf{3}_\mathrm{R}) \,,
\end{split}\end{equation} 
as well as two sextets $\Delta_L\sim(\mathbf{6}_L,\mathbf{1}_R)$,
$\Delta_R\sim(\mathbf{1}_L,\mathbf{6}_R)$ with components
\begin{equation}
\Delta_{L,R}=\left(
 \begin{array}{ccc}
 \Delta^{0}_{11} & \frac{\Delta^{-}_{12}}{\sqrt{2}} & \frac{\Delta^{q}_{13}}{\sqrt{2}}\\ 
 \frac{\Delta^{-}_{12}}{\sqrt{2}} &  \Delta^{--}_{22} & \frac{\Delta^{q-1}_{23}}{\sqrt{2}}\\
 \frac{\Delta^{q}_{13}}{\sqrt{2}} & \frac{\Delta^{q-1}_{23}}{\sqrt{2}} & \Delta^{2q}_{33}
 \end{array}
\right)_{L,R} \,.
\end{equation}
The above fields transform as $\phi\to U_L\phi\, U_R^{\dagger}$,
$\rho\to U_L\rho\, U_R^{T}$, $\Delta_L\to U_L\Delta_L\, U_L^{T}$ and
$\Delta_R\to U_R\Delta_R\, U_R^{T}$ under
$\mathrm{SU(3)_L\otimes SU(3)_R}$.
The symmetry breaking pattern in the scalar sector is assumed to be driven by
\begin{equation}\begin{split}
\vev{\phi}=\frac{1}{\sqrt{2}}\text{diag}(k_1 ,k_2 ,n)\hspace{2mm} \vev{\Delta_L}&=\frac{1}{\sqrt{2}}\text{diag}(v_L , 0, 0),\hspace{2mm} \vev{\Delta_R}=\frac{1}{\sqrt{2}}\text{diag}(v_R , 0, 0)\,,\\&
\vev{\rho}=\left(\begin{array}{ccc}
0  & k_3 & 0 \\
0  & 0 &0  \\
 0 & 0 &  0\\
\end{array}\right)\,,
\end{split}\end{equation}
where the vacuum expectation values (VEV) $n$ and $v_R$ set the scale
of symmetry breaking down to the standard model one, and subsequently
$k_1$, $k_2$, $k_3$ and $v_{L}$ are responsible for the SM electroweak
spontaneous symmetry breaking. Thus, for consistency,
$n\,,v_R\gg k_1\,,k_2\,,k_3\,,v_L$. In what follows we will explore
spontaneous symmetry breaking patterns determined by the value of
$n/v_R$, as well as the natural expected hierarchy for the remaining
VEVs $k_1\,,k_2\,,k_3\,,v_L$.

\section{Particle masses}  
\label{sec:particle-masses}

We now turn to the Yukawa interactions of the theory. These are
similar to the ones present in the most popular left-right symmetric
models, namely
\begin{equation}
\label{lagyuk}
\begin{split}\mathcal{L}_{y}=&\sum _{\alpha,\beta=1}^{2}\left(h^{Q}_{\alpha\beta}\overline{Q}_{L}^{\,\alpha}\phi^* Q_{R}^{\beta}\right)+\sum _{\alpha=1}^{2}\left(h^{Q}_{\alpha 3}\overline{Q}_{L}^{\,\alpha}\rho^* Q_{R}^{3}+h^{Q}_{3\alpha}\overline{Q}^{\,3}_{L}\rho Q^\alpha_{R}\right)+h^{Q}_{33}\overline{Q}^{\,3}_{L}\phi Q^3_{R}\\&+ \sum_{a,b=1}^{3} \left[ h^{\ell}_{ab}\overline{\psi}_{L}^{\,a}\phi \psi_{R}^{b}+f_{ab}\left(\overline{\psi}_{L}^{\,c\, a}\Delta_{L}^\dagger\psi_{L}^{b}+\overline{\psi}_{R}^{\,c\, a}\Delta_{R}^\dagger\psi_{R}^{b}\right)\right]+\mathrm{h.c.}
\end{split}
\end{equation}
with $h^Q=(h^Q)^{\dagger}$ and $h^{\ell}=(h^{\ell})^\dagger$.  After
spontaneous symmetry breaking, the first line in Eq.~(\ref{lagyuk})
produces the following Dirac mass matrices for the Standard Model and
exotic quarks:
\begin{equation}\label{mass_up_down}
M^{u}=\frac{1}{\sqrt{2}}\left(
\begin{array}{ccc}
h_{11}^Qk_2 &  h_{12}^Qk_2 & 0 \\
 h_{21}^Qk_2 &  h_{22}^Qk_2 &  0 \\
-h_{31}^Q k_3 &  -h_{32}^Q k_3 &  h_{33}^Qk_1 \\
\end{array}
\right)\,,\qquad 
M^{d}=\frac{1}{\sqrt{2}}\left(
\begin{array}{ccc}
 h_{11}^Qk_1 &  h_{12}^Qk_1 & h_{13}^Q k_3 \\
 h_{21}^Qk_1 &  h_{22}^Qk_1 & h_{23}^Q k_3 \\
0 & 0 &  h_{33}^Qk_2 \\
\end{array}
\right)\,,
\end{equation}
\begin{equation}\label{mass_exotic}
M^{J^{-q-\frac{1}{3}}}=\frac{1}{\sqrt{2}}\left(
\begin{array}{cc}
 h_{11}^Qn &  h_{12}^Qn \\
 h_{21}^Qn &  h_{22}^Qn \\
\end{array}
\right)\,,\qquad  M^{J^{q+\frac{2}{3}}}=h_{33}^Qn.
\end{equation}
Notice that in the absence of $\rho$, the quark mass matrices are
block diagonal and the ratio between the electroweak scales $k_1$ and
$k_2$ is fixed by the bottom and top masses
$k_2/k_1=m_b/m_t$. Moreover, the upper 2$\times$2 blocks in $M^{u}$
and $M^{d}$ are proportional. This implies that, in this limit the CKM
matrix is trivial. As a result in our model $\rho$ generates all
entries of the quark mixing matrix, the Cabibbo
angle, $V_{ub}$ as well as $V_{cb}$. \\

For leptons the situation is qualitatively different. The charged
lepton mass matrix is simply given by
\begin{equation}
m_{ab}^\ell=\frac{k_2}{\sqrt{2}}h^\ell_{ab}\hspace{1mm},
\end{equation}
and the new leptons $\chi^q_{L,R}$ form heavy Dirac pairs with masses
\begin{equation}
m_{ab}^\chi=\frac{n}{\sqrt{2}}h^\ell_{ab}\hspace{1mm}.
\end{equation}
Concerning neutrinos, their mass matrix can be written as
\begin{equation}
m_{\nu}=\left(\begin{array}{ccc}
M_L  & m_{D} \\
m_{D}^T & M_R \\
\end{array}\right)
\hspace{0.5mm},
\end{equation}
where
\begin{equation}
M_L=2f_{ab}v_L,\hspace{2mm}m_{D}=h^\ell_{ab}k_1, \hspace{2mm}M_R=2f_{ab}v_R\hspace{0.3mm}.
\end{equation}
Thus we obtain a combination of type I and type II seesaw mechanisms,
the same situation than in the $\mathrm{SU(2)_{L}\otimes SU(2)_{R}}$
models:
\begin{equation}
m_1\approx M_L -m_{D} M_R^{-1} m_{D}^T\,,\qquad m_2 \approx M_R\,.
\end{equation}

We now turn to the gauge boson masses which come as usual from their
couplings with the scalars present in the theory. The relevant
covariant derivative is defined as
\begin{equation}
D_\mu=\partial_\mu-i\frac{g}{2}\textbf{W}^L_\mu-i\frac{g}{2}\textbf{W}^R_\mu-ig_X X B_{\mu}\,.
\end{equation}
where the vector bosons are expressed as a matrix
\begin{equation}
\textbf{W}^{L,R}_\mu=\sum^8_{i=1}W^i_{L\hspace{0.3mm}\mu}\Lambda_i=
\left (\begin{array}{ccc}
W^3+\frac{1}{\sqrt{3}}W^8 & W^+ & V^{-q}\\
W^- & -W^3+\frac{1}{\sqrt{3}}W^8 & V^{\prime \hspace{0.1mm}-q-1}\\
V^{q} & V^{\prime \hspace{0.1mm}q+1} & -\frac{2}{\sqrt{3}}W^8\\
\end{array}\right )_{L,R}\hspace{1mm},
\end{equation}
with $\Lambda_i$ as the Gell-Mann matrices. There are in total 17 gauge bosons in the physical basis, the photon:
$\gamma$, four electrically neutral states: $Z_L$, $Z_R$,
$Z^\prime_L$, $Z^\prime_R$, four charged bosons: $W^{\pm}_L$,
$W^{\pm}_R$, four states with charge $q+1$: $X_{L}^{\pm(1+q)}$,
$X_{R}^{\pm(1+q)}$, and four with charge $q$: $Y_{L}^{\pm q}$,
$Y_{R}^{\pm q}$. One can determine the mass matrices and diagonalize
them (assuming the VEV hierarchy $v_L\ll k_{1,2,3}\ll v_R,n$) to obtain
the gauge boson masses
\begin{equation}\begin{split}
m_{W_L}^2\approx&\frac{g^2}{2}(k_1^2+k_2^2+k_3^2+2v_L^2)\hspace{1mm},\\
m_{Z_L}^2\approx&\frac{g^2}{2\cos ^2\theta_W}(k_1^2+k_2^2+k_3^2+4v_L^2)\hspace{1mm},\\
m_{Z_L^\prime}^{2}\sim & \hspace{0.7mm}O(n^2)\hspace{1mm},\\
m_{W_R}^2\approx&\frac{g^2}{2}(k_1^2+k_2^2+k_3^2+2v_R^2)\hspace{1mm},\\
m_{Z_R}^2\approx&\frac{g^2}{2\cos ^2\theta_W}(k_1^2+k_2^2+k_3^2+4v_R^2)\hspace{1mm},\\
m_{Z_R^\prime}^{ 2}\sim & \hspace{0.7mm} {\cal O}(n^2+v_R^2)\hspace{1mm},\\
m_{X_L}^2\approx & m_{Y_L}^2\approx m_{Y_R}^2\sim  {\cal O}(n^2)\hspace{1mm},\\
 m_{X_R}^2\sim & {\cal O}(n^2+v_R^2)\hspace{1mm}.
\end{split}\end{equation}

\section{Scalar potential}

The most general CP conserving scalar potential compatible with all
the symmetries of the model is
\begin{equation}
V=V_\Delta+V_\phi+ V_\rho +V_{\text{mix}},
\end{equation}
with
\begin{equation}\begin{split}
&V_\Delta=\mu_\Delta^2 \left[ \Tr(\Delta_L\Delta_L^{\dagger})+ \Tr(\Delta_R\Delta_R^{\dagger})\right]+\lambda_1\left[  \Tr(\Delta_L\Delta_L^{\dagger}) ^2+  \Tr(\Delta_R\Delta_R^{\dagger}) ^2\right]\\&\qquad+\lambda_2\left[ \Tr(\Delta_L\Delta_L^{\dagger}\Delta_L\Delta_L^{\dagger})+ \Tr(\Delta_R\Delta_R^{\dagger}\Delta_R\Delta_R^{\dagger})\right]\hspace{1mm},
\\
&V_\phi=\mu_\phi^2 \Tr(\phi\phi^{\dagger})+\lambda_3 \Tr(\phi\phi^{\dagger})^2+\lambda_4  \Tr(\phi\phi^{\dagger}\phi\phi^{\dagger})+f_\phi(\phi\phi\phi+ \mathrm{h.c.})\hspace{1mm},
\\
& V_\rho=\mu_\rho^2 \Tr(\rho\rho^{\dagger})+\lambda_5 \Tr(\rho\rho^{\dagger})^2+\lambda_6  \Tr(\rho\rho^{\dagger}\rho\rho^{\dagger})\hspace{1mm},
\\
&V_{\text{mix}}=\lambda_7 \Tr(\phi\phi^\dagger)\Tr(\rho\rho^\dagger)+\lambda_8 \left[\Tr(\phi\phi^\dagger\rho\rho^\dagger)+\Tr(\phi^\dagger\phi\rho^T\rho^*)\right]+\lambda_9\Tr(\Delta_R\Delta_R^\dagger)\Tr(\Delta_L\Delta_L^\dagger)\hspace{1mm},
\\
&\qquad +\lambda_{10}\Tr(\phi\phi^\dagger)\left[ \Tr(\Delta_R\Delta_R^\dagger)+ \Tr(\Delta_L\Delta_L^\dagger)\right]
+\lambda_{11}\left[ \Tr(\phi^\dagger\phi\Delta_R\Delta_R^\dagger)+ \Tr(\phi\phi^\dagger\Delta_L\Delta_L^\dagger)\right]
\\
&\qquad 
+\lambda_{12}\left[ \Tr(\phi\Delta_R\phi^T\Delta_L^*)+\mathrm{h.c.}\right]+\lambda_{13}\Tr(\rho\rho^\dagger)\left[ \Tr(\Delta_R\Delta_R^\dagger)+ \Tr(\Delta_L\Delta_L^\dagger)\right]
\\&\qquad+\lambda_{14}\left[ \Tr(\rho^T\rho^*\Delta_R\Delta_R^\dagger)+ \Tr(\rho\rho^\dagger\Delta_L\Delta_L^\dagger)\right]+\lambda_{15}\Tr(\phi\rho^T\phi^*\rho^\dagger)\hspace{1mm}.
\end{split}\end{equation}

\begin{figure}
  \centering
    \includegraphics[width=0.5\textwidth]{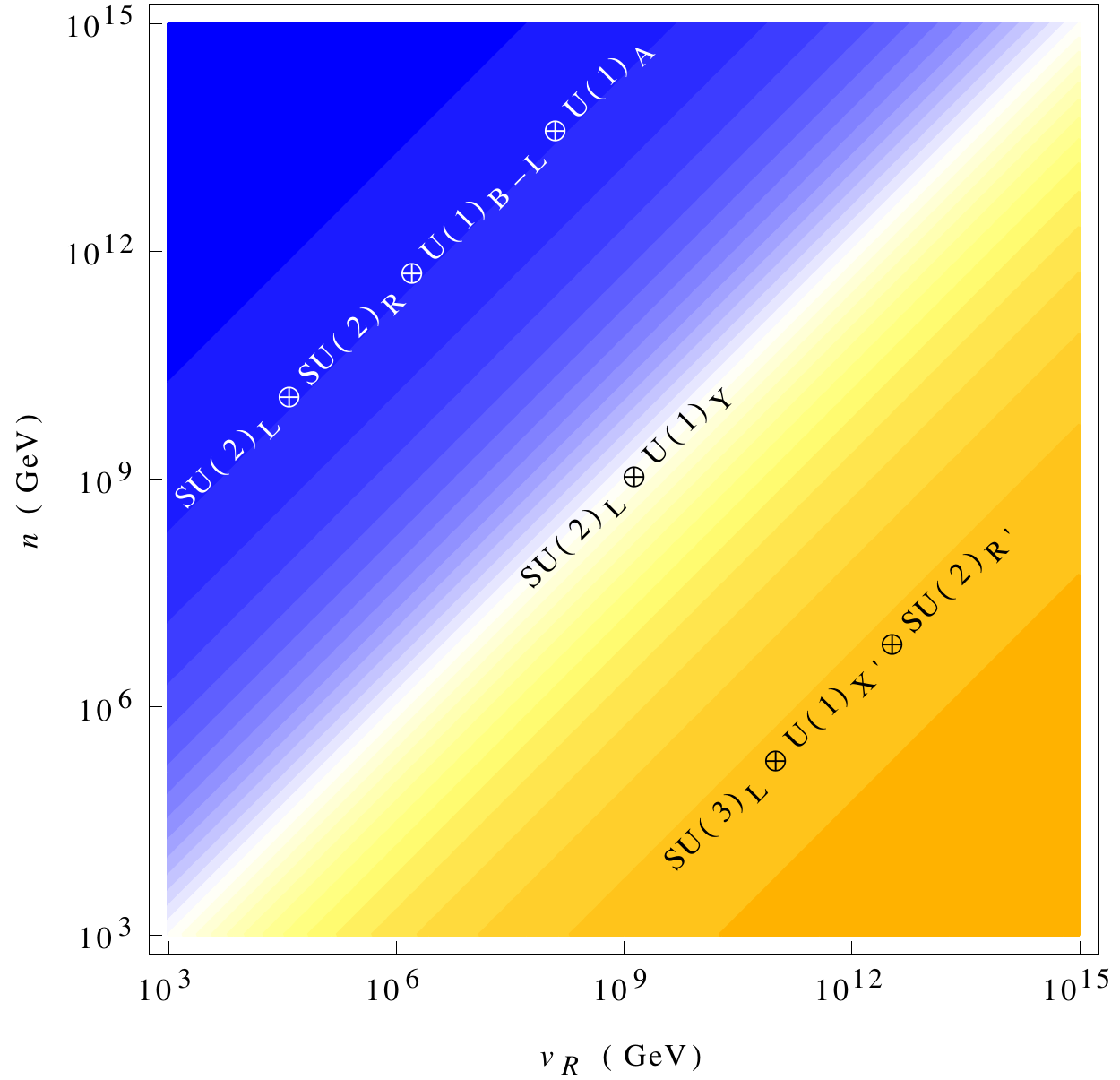}
    \caption{ Phase diagram of the \TrTrTrOne electroweak theory,
      discussed in Sec.~\ref{sec:spont-symm-break}, see also
      Fig.~\ref{ssbpath}. The \sm singlet VEVs $n$ and $v_R$ are in
      GeV units and their ratio is determined by
      Eq.~(\ref{eq:paths}). }
  \label{ssb-phases}
\end{figure}

The extremum conditions can be solved in terms of the dimensionful
parameters of the potential
\begin{equation}
\begin{split}
&\mu_\Delta^2=-\frac{1}{2}\left[2v_R^2(\lambda_1+\lambda_2)+v_L^2\lambda_9+(n^2+k_1^2+k_2^2)\lambda_{10}+k_1^2\lambda_{11}+\frac{v_L}{v_R}k_1^2\lambda_{12}+k_3^2\lambda_{13}\right]\hspace{1mm},\\&
\mu^2_\phi=-\frac{1}{2}\left[(v_L^2+v_R^2)\lambda_{10}+2(n^2+k_1^2+k_2^2)\lambda_3+(n^2+2k_2^2)\lambda_4+k_3^2\lambda_7-\frac{k_2^2k_3^2}{n^2-k_2^2}\lambda_8\right]\hspace{1mm},\\&
\mu^2_\rho=-\frac{1}{2}\left[2k_3^2(\lambda_5+\lambda_6)+(n^2+k_1^2+k_2^2)\lambda_7+(k_1^2+k_2^2)\lambda_8+(v_R^2+v_L^2)\lambda_{13}+v_L^2\lambda_{14}\right]\hspace{1mm},\\&
f_{\phi}=\frac{nk_2\left[2(n^2-k_2^2)\lambda_4-k_3^2\lambda_8\right]}{6\sqrt{2}k_1(n^2-k_2^2)}\hspace{1mm},
\end{split}\end{equation}
together with the conditions: 
\begin{equation}\begin{split}
&v_Lv_R\left[2(\lambda_1+\lambda_2)-\lambda_{9}-\frac{k_3^2}{v_R^2-v_L^2} \lambda_{14}\right]-\lambda_{12}k_1^2=0\hspace{1mm},\\
&\frac{n^2-k_2^2}{k_1^2}(k_1^2-k_2^2)\lambda_4-\frac{\lambda_{11}}{2}(v_L^2+v_R^2)-\lambda_{12}v_Lv_R-\frac{n^2k_3^2(k_1^2-k_2^2)}{2k_1^2(n^2-k_2^2)}\lambda_8=0\hspace{1mm},\\
&\lambda_{15}=0\,.
\end{split}\end{equation}
Assuming $\epsilon\equiv\frac{k_3^2}{v_R^2-v_L^2}\ll1$ and natural
values for the quartic couplings, the first condition leads to the
well-known VEV seesaw relation
\begin{equation}
\label{eq:vevseesaw}
v_Lv_R=\frac{\lambda_{12}k_1^2}{2(\lambda_1+\lambda_2)-\lambda_{9}-\epsilon \lambda_{14}}\hspace{1mm},
\end{equation}
which characterizes dynamically the seesaw mechanism. This is
consistent with the hierarchy between the VEVs $v_R\gg k_1 \gg v_L$
and consequently, the second condition reduces to
\begin{equation}
\label{eq:paths}
\lambda_4 n^2 \frac{(k_1^2-k_2^2)}{k_1^2}\approx\frac{\lambda_{11}}{2}v_R^2
\end{equation}
at leading order in the \sm singlet VEVs $n$ and $v_R$. The latter
shows clearly, as seen in Fig.~\ref{ssb-phases}, that the primordial
\TrTrTrOne theory can break either directly to the \sm (central part
of the plot) or through the intermediate  \lr or $\mathrm{SU(3)_c \otimes SU(3)_L \otimes U(1)}$ phases,
corresponding to the upper and lower regions, respectively. This
behavior is mainly controlled by the quartic parameters $\lambda_{4}$
and $\lambda_{11}$ in the scalar potential. More details in the next
section.

\section{Spontaneous symmetry breaking}
\label{sec:spont-symm-break}

\begin{figure}
  \centering
    \includegraphics[width=0.89\textwidth]{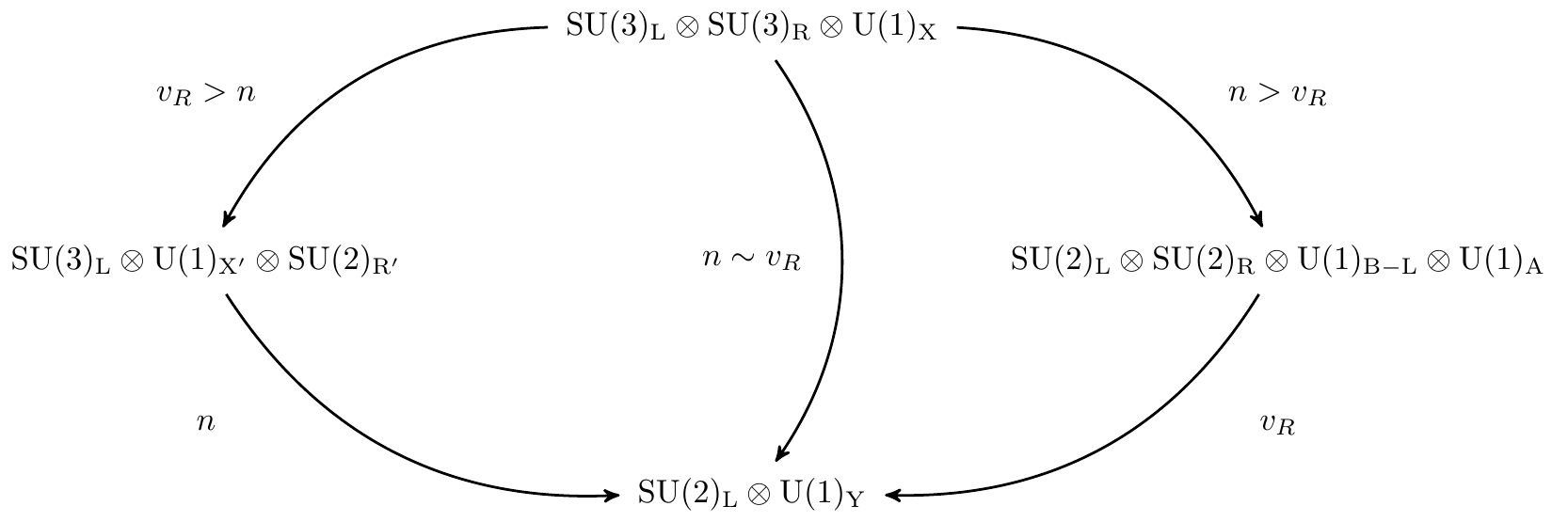}
    \caption{ Spontaneous symmetry breaking paths in the \TrTrTrOne
      electroweak theory. See also Fig.~\ref{ssb-phases} and the VEV
      seesaw relation in Eq.~(\ref{eq:vevseesaw}) as well as the
      breaking pattern determining condition in Eq.~(\ref{eq:paths}).}
  \label{ssbpath}
\end{figure}

In order to recover the Standard Model at low energies we need to
break the \TrTrTrOne symmetry.  The breaking of the gauge structure
can be achieved in several ways (see Fig.~\ref{ssbpath}) depending on
the value of $n/v_R$. For $n>v_R \gg k_{1,2,3}$ the symmetry breaking
pattern is:
\begin{equation}
\mathrm{SU(3)_L\otimes SU(3)_R\otimes U(1)_{X}}\xrightarrow{n} \mathrm{SU(2)_L\otimes SU(2)_R\otimes U(1)_{B-L}\otimes U(1)_{A}}\xrightarrow{v_R}\mathrm{SU(2)_L\otimes U(1)_{Y}}\hspace{1mm}.
\end{equation}
At this stage one has
$\vev{\phi}=\frac{1}{\sqrt{2}}\text{diag}(0 ,0 ,n)$ breaking the
$T^8_{L, R}$ generators but preserving $T^8_{L}+T^8_{R}$, and since $\phi$ carries no
$X$ charge, the resulting gauge group is
$\mathrm{SU(2)_L\otimes SU(2)_R\otimes U(1)_{B-L}\otimes U(1)_{A}}$, with
\begin{equation}\begin{split}
&\frac{B-L}{2}=\beta (T^8_L+T^8_R)+X\,,\\&
A=\beta (T^8_L+T^8_R)-X\,.
\end{split}\end{equation}
It is important to notice that $A$ is not involved in the electric charge definition since it reads
\begin{equation}
Q=T^3_R+T^3_L+\beta (T^8_L+T^8_R)+X\equiv T^3_R+T^3_L+\frac{B-L}{2}\,.
\end{equation}
The potential acquires the
form
$V=V(\phi^\prime,\rho^\prime,\Delta_R^\prime,\Delta_L^\prime,\dots)$
where 
\begin{equation}\begin{split}
\phi^\prime &\sim(\two_\mathrm{L} ,\two^*_\mathrm{R} ,0,0)\hspace{1mm},\qquad
\rho^\prime\sim(\two_\mathrm{L} ,\two_\mathrm{R} ,0,-\frac{4q+2}{3})\hspace{1mm},\\
\Delta_L^\prime &\sim(\three_\mathrm{L} ,\one_\mathrm{R} ,-1,\frac{1-4q}{3})\hspace{1mm},\qquad
\Delta_R^\prime\sim(\one_\mathrm{L} ,\three_\mathrm{R} ,-1,\frac{1-4q}{3})\hspace{1mm},\\
\end{split}\end{equation}
are the $2\times 2$ upper left sub-matrices contained in the original
\TrTrTrOne scalar multiplets in notation $\mathrm{(SU(2)_L,SU(2)_R,\frac{B-L}{2},A)}$, and the dots stand for the extra scalars.

Apart from the existence of the scalar multiplet $\rho$ and the extra symmetry $\mathrm{U(1)_{A}}$ the
situation here resembles the popular \lr electroweak model in
\cite{mohapatra:1980ia}. The second step of the spontaneous symmetry
breaking is carried out by
$\vev{\Delta_R}=\frac{1}{\sqrt{2}}\text{diag}(v_R ,0 ,0)$. We note here that the generators $T_R^3$ and $A$ are broken at
this stage, though the combination
\begin{equation}
T^3_R+\frac{B-L}{2}=Y\hspace{1mm},
\end{equation}
remains unbroken and hence can be identified with the \sm hypercharge.
In this scenario, an $\mathrm{SU(2)_L\otimes SU(2)_R\otimes U(1)_{B-L}}$
structure emerges as the effective theory at lower scales.
Moreover, in this case, at energy scales above $v_R$, we expect to
observe new physics associated to a 331 model, such as virtual
effects associated with exotic quarks and leptons and new gauge
bosons, even if these particles are too heavy to show up directly.\\

Alternatively, if the VEV hierarchy is $v_R>n \gg k_{1,2,3}$, the
symmetry breakdown follows a different route\footnote{Notice that this
  extra $SU(2)_R^\prime$ group comes from the fact we keep $\beta$
  arbitrary: it is easy to see that one can recover the usual
  331~\cite{Singer:1980sw} electroweak group for q=0, which allows
  $\Delta_{33}$ to have a non-zero vev.}
\begin{equation}
\mathrm{SU(3)_L\otimes SU(3)_R\otimes U(1)_{X}}\xrightarrow{v_R} \mathrm{SU(3)_L\otimes U(1)_{X^\prime}\otimes SU(2)_{R'}}\xrightarrow{n}\mathrm{SU(2)_L\otimes U(1)_{Y}}\hspace{1mm}.
\end{equation}
Now $\vev{\Delta_R}=\frac{1}{\sqrt{2}}\text{diag}(v_R ,0 ,0)$ breaks the $\mathrm{SU(3)_R}$ group down to $\mathrm{SU(2)_{R'}}$, generated by $\{T^6_{R},\,T^7_{R},\,\frac{1}{2}(\sqrt{3}T^8_{R}-T^3_{R})\}$. Simultaneously, $\mathrm{U(1)_{X}}$ is broken by $v_R$ but the combination $X'=\frac{\beta+\sqrt{3}}{4}(T^8_R+\sqrt{3}T^3_{R})+X$ is preserved so our
theory becomes an effective 331 model with an additional $\mathrm{SU(2)_{R'}}$ symmetry at intermediate energies. In terms of the generators of the intermediate symmetries, electric charge reads 
\begin{equation}
Q=T^3_L+T^3_R+\beta (T^8_L+T^8_R)+X\equiv T^3_L+\beta T^8_L	+\frac{\sqrt{3}\beta-1}{4}(\sqrt{3}T^8_{R}-T^3_{R})+X'\,.
\end{equation} The
potential in this case can be written as
$V=V(\phi^A,\phi^{B},\rho^A,\rho^{B},\Delta_L^\prime,\dots)$,
where the relevant fields transform under
$\mathrm{(SU(3)_L,SU(2)_{R'},X')}$ as
\begin{equation}\begin{split}
\phi^A&\sim(\three_\mathrm{L},\one_\mathrm{R'},\frac{q-1}{3})\hspace{1mm},\qquad
\phi^B\sim(\three_\mathrm{L}, \two_\mathrm{R'},\frac{1-q}{6})\hspace{1mm},\\
\rho^A&\sim(\three_\mathrm{L},\one_\mathrm{R'}, \frac{q+2}{3})\hspace{1mm},\qquad
\rho^B\sim(\three_\mathrm{L},\two_\mathrm{R'},\frac{5q+1}{6})\hspace{1mm},\\
\Delta_L^\prime &\sim(\six_\mathrm{L},\one_\mathrm{R'},\frac{2(q-1)}{3})\hspace{1mm}.\\
\end{split}\end{equation}
In a second step the triplet $\phi^A=\frac{1}{\sqrt{2}}(0 ,0 ,n)^T$
breaks the $\mathrm{SU(3)_L\otimes SU(2)_{R'}\otimes U(1)_{X^\prime}}$ symmetry down to
the Standard Model.  For this VEV hierarchy, signals associated to
exotic fermions are expected at intermediate energies, while new
physics related to left-right symmetry, like neutrino masses, emerges
at higher energies.\\

Finally, a third situation in which $n\sim v_R \gg k_{1,2,3}$ is also
possible. The \TrTrTrOne gauge group in this case is broken directly
to the that of the \sm :
\begin{equation}
\mathrm{SU(3)_L\otimes SU(3)_R\otimes U(1)_{X}}\xrightarrow{n,\,v_R}\mathrm{SU(2)_L\otimes U(1)_{Y}}\hspace{1mm}.
\end{equation}
In this scenario one expects new physics associated to left-right and
331 symmetries at comparable energy scales.

\section{Discussion and conclusion}
\label{sec:disc-concl}

We have proposed a fully realistic scheme based on the \TrTrTrOne
gauge group. Quantum consistency requires that the number of families
must match the number of colors, hence predicting the number of
generations. In the simplest realization neutrino masses arise from a
dynamical seesaw mechanism in which the smallness of neutrino masses
is correlated with the observed V-A nature of the weak
interaction. Depending on the symmetry breaking path to the \sm (see
Figs.~\ref{ssb-phases} and \ref{ssbpath}) one recovers either a \lr
theory or one based on the $\mathrm{SU(3)_c \otimes SU(3)_L \otimes U(1)_{X'}}$ gauge symmetry.
This illustrates the versatility of the theory since, depending on a
rather simple input parameter combination, it can mimic either of two
apparently irreconcilable pictures of nature: one based upon
left-right symmetry and another characterized by the \TrTrOne gauge
group. Either of these could be the ``next'' step in the quest for new
physics.

\begin{appendix}
 \section{Anomaly cancellation in \TrTrTrOne}
\label{AnomalyCancellation}

In this section we outline the anomaly cancellation in the \TrTrTrOne
model. The potential anomalies  are
$[\mathrm{SU(3)_c}]^2\otimes \mathrm{U(1)_{X}},\hspace{1mm}
[\mathrm{SU(3)_L}]^3,\hspace{1mm}[\mathrm{SU(3)_R}]^3,\hspace{1mm}[\mathrm{SU(3)_L}]^2\otimes
\mathrm{U(1)_{X}},\hspace{1mm}[\mathrm{SU(3)_R}]^2\otimes
\mathrm{U(1)_{X}},\hspace{1mm}[\mathrm{Grav}]^2\otimes
\mathrm{U(1)_{X}},\hspace{1mm}[\mathrm{U(1)_{X}}]^3$.
First notice that the $[\mathrm{SU(3)_c}]^2\otimes \mathrm{U(1)_{X}}$ anomaly cancels
straightforwardly
\begin{equation}
\sum_{\text{quarks}}X_{qL}-\sum_{\text{quarks}}X_{qR}= \left ( -\frac{q}{3}\right ) \times 3\times 3\times 2+\left ( \frac{q+1}{3}\right )\times 3\times 3- \left (-\frac{q}{3}\right ) \times 3\times 3\times 2-\left ( \frac{q+1}{3}\right )\times 3\times 3=0\hspace{1mm}.
\end{equation}
Since triplets and anti-triplets contribute with opposite sign to the
$[\mathrm{SU(3)_L}]^3$ and $[\mathrm{SU(3)_R}]^3$ anomalies, these are only canceled if
the number of triplets is equal to the number of anti-triplets, which
is the case in our model. Thus from $[\mathrm{SU(3)_R}]^3$ and $[\mathrm{SU(3)_L}]^3$ we
conclude that if $f$ is the number of lepton families, $n$ is the
number of quark triplets and $m$ is the number of quark anti-triplets
the following equation must hold:
\begin{equation}
3\times f+3\times 3\times n=3\times 3\times m \to f=3 (m-n)\hspace{1mm}.
\end{equation}

Next, we consider $[\mathrm{SU(3)_L}]^2\otimes \mathrm{U(1)_{X}}$ and
$[\mathrm{SU(3)_R}]^2\otimes \mathrm{U(1)_{X}}$ which are canceled independently
because the sum of all $X$ charges is equal to zero:
\begin{equation}
\label{anomaly1}
\begin{split}
\sum_{\text{fermions}}X_L =\left ( \frac{q-1}{3}\right ) \times 3\times 3+\left ( -\frac{q}{3}\right ) \times 3\times 3\times 2+\left ( \frac{q+1}{3}\right )\times 3\times 3=0\hspace{1mm},\\
\sum_{\text{fermions}}X_R =\left ( \frac{q-1}{3}\right ) \times 3\times 3+\left ( -\frac{q}{3}\right ) \times 3\times 3\times 2+\left (  \frac{q+1}{3}\right )\times 3\times 3=0\hspace{1mm}.
\end{split}
\end{equation}                   
In more general terms, this condition is the equivalent to
\begin{equation}
\sum_{\text{fermions}}X_L =\left (\frac{q-1}{3}\right ) \times 3\times f+\left ( -\frac{q}{3}\right ) \times 3\times 3\times m+\left ( \frac{q+1}{3}\right )\times 3\times 3 \times n=0\hspace{1mm}.
\end{equation}
Plugging the previous result $f=3(m-n)$ into the above relation
implies that $m=2n$ and $f=3n$, independently of the value of
$q$. This condition is complemented by QCD asymptotic freedom, that
requires the number of quark flavors to be less than or equal to 16
leading to $3\times (m+n)=9 n\leq 16$, leaving only one positive
integer solution $n=1$, hence we must have $f=3$ generations.

Finally, the theory is free from gravitational anomaly 
$[\mathrm{Grav}]^2\otimes \mathrm{U(1)_{X}}$ if
\begin{equation}
\sum_{\text{fermions}}X_L =\sum_{\text{fermions}}X_R,
\end{equation}
which is trivially satisfied, while the cancellation of the
$[\mathrm{U(1)_{X}}]^3$ anomaly follows from
\begin{equation}
\begin{split}
\sum_{f_L ,q_L}X_L^3 -\sum_{f_R ,q_R}X_R^3&=\left ( \frac{q-1}{3}\right )^3 \times 3\times 3+\left (  -\frac{q}{3}\right )^3 \times 3\times 3\times 2+\left ( \frac{q+1}{3}\right )^3\times 3\times 3\\&
-\left ( \frac{q-1}{3}\right )^3 \times 3\times 3-\left (  -\frac{q}{3}\right )^3 \times 3\times 3\times 2-\left ( \frac{q+1}{3}\right )^3\times 3\times 3=0
\end{split}
\end{equation}
Notice that L-R symmetry plays an important role in anomaly
cancellation since it automatically implies that the fermion content
of the model satisfy $X_L=X_R$ for every multiplet and all particles
are arranged in chiral multiplets. We also remark that in our
\TrTrTrOne theory the anomaly cancellation holds irrespective of the
value of the $\beta$ parameter of the electric charge generator in
Eq.~(\ref{elcharge}).

\end{appendix}

\section{Acknowledgements}

We thank Renato Fonseca, Martin Hirsch, 
P.V. Dong and D.T. Huong for useful discussions.
Work supported by Spanish grants FPA2014-58183-P, Multidark
CSD2009-00064, SEV-2014-0398 (MINECO) and PROMETEOII/2014/084
(Generalitat Valenciana).  C.A.V-A. acknowledges support from CONACyT
(Mexico), grant 274397.  M.~R. was supported by JAEINT-16-00831.

\bibliographystyle{utphys}
\bibliography{3331,merged_Valle,newrefs}
 
\end{document}